\documentclass[prl,showpacs,showkeys,twocolumn,10pt]{revtex4-1}
\usepackage{amsfonts,bbold,amsmath,amssymb,graphicx,epstopdf,verbatim,dsfont,color}
\usepackage[english]{babel}

\begin{document}

\title{Dynamic polaron response from variational imaginary time evolution}
\author{Dries Sels}
\affiliation{Department of Physics, Boston University, Boston, MA 02215, USA}
\affiliation{TQC, Universiteit Antwerpen, B-2610 Antwerpen, Belgium}
\date{\today}

\begin{abstract}
An variational expression for the zero temperature polaron impedance is obtained by minimizing the free energy in a generalized quadratic Feynman model. The impedance function of the quadratic model serves as the variational parameter. It is shown that a very small change in the energy can be accompanied by a large change in the optical conductivity. This is related to the insensitivity of the Jensen-Feynman free energy to the UV properties of the model. Analytic and numeric results are derived for the Fr\"ohlich polaron in weak and strong coupling. Standard results are recovered at weak coupling but, more importantly, strong coupling inconsistencies are removed.
\end{abstract}
\maketitle
Since the seminal work of Landau on the motion of electrons in solids~\cite{Landau}, the polaron problem has been of considerable theoretical interest~\cite{Alexandrov}, mostly because it is one of the simplest examples of a particle interacting with a field. Over the years it has been a testing ground for various theoretical techniques in quantum field theory, from Diagrammatic quantum Monte Carlo~\cite{DQMC} to the recent application of renormalziation group theory~\cite{Grusdt}. 

Recent experimental advances in the field of ultra-cold atoms offer a new platform for the experimental study of polarons~\cite{Grusdt, Timmermans, Tempere, Casteels}. By immersing an impurity atom in a BEC one can study the formation of polarons where the condensates Bogoliubov excitations serve as the scalar field. Feshbach resonances make it possible to realize tunable interactions between the impurity and host atoms. Whether these systems truly behave as Fr\"ohlich polarons remains an open question~\cite{Grusdt2}. Due to the experimental relevance for cold atom experiments there has been a revived interest in the non-equilibrium behaviour of polarons. 

Despite numerous works devoted to the Fr\"ohlich polaron, its dynamics is far from completely understood over the full range of interactions. Even close to equilibrium properties such as the optical absorption and mobility remain areas of interest. In particular, the discrepancy between the Kadanoff~\cite{Kadanoff} and the Feynman-Hellwarth-Iddings-Platzman~\cite{FHIP} (FHIP) mobility was only properly understood after a recent derivation of the mobility using Wigner distribution functions~\cite{SBmobility,SBBoltzman}. Remarkably neither the path integral approach of FHIP nor the relaxation time approximation of Kadanoff give the correct result. Instead the mobility was found to agree with that of Los~\cite{Los}. For a detailed overview I refer to Ref.~\cite{DevreeseVarenna}.

Path integrals offer unique insight in the polaron problem because the bosonic degrees of freedom can be eliminated exactly. However, in order to proceed one must make approximations and there is a surprising gap between the accuracy of equilibrium and non-equilibrium methods. Whereas Feynman's variational results~\cite{Feynman} are excellent for the free energy over the complete interaction range, it's dynamical predictions are rather poor. Neither the weak coupling mobility of~\cite{FHIP} nor the strong coupling absorption obtained in~\cite{DSG} are accurate. Moreover, it was shown by Devreese \emph{et. al.} in~\cite{DSG} that the linewidth at strong coupling is in violation with the uncertainty principle~\cite{KED}.
\begin{figure}[h]
\centering
\includegraphics[width=0.45\textwidth]{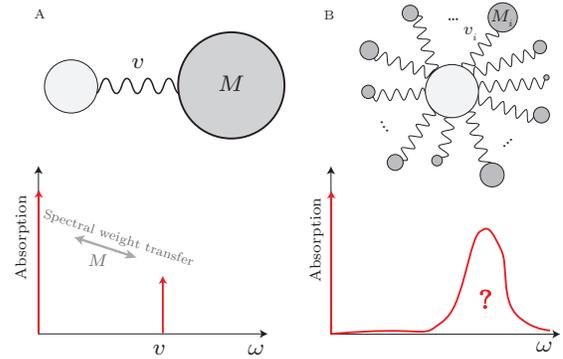}
\caption{(A) The optical absorption of Feynman's variational model, as parametrized by the effective mass $M$ and the frequency $v$. The model treats the polaron as two masses connected by a spring. (B) Its generalization to many degrees of freedom allows the construction of any absorption spectrum.}
\label{fig:ModelAbsorption}
\end{figure}

In this note I will address this issue in detail. The inaccuracy is related to an important assumption in~\cite{FHIP}. The entire procedure to calculate the mobility at low temperatures is based on \emph{the supposition that the model Lagrangian, which gives a good fit to the ground-state energy at zero temperature, will also give the dynamical behaviour of the system}~\cite{FHIP}. One should note however that the dynamical behaviour of Feynman's model is completely fixed \emph{a priori}, as it simply consists of two harmonic degrees of freedom. The preserve translattional invariance one eigenmode is always at zero frequency. The frequency of the other mode and the relative spectral weight in the peaks (or effective mass) are the free parameters, see Fig.~\ref{fig:ModelAbsorption}~A. Since this spectrum deviates significantly from the real spectrum of the polaron, the assumption in FHIP is inadequate. This was already recognized by FHIP, who state that the dissipative part of the model must only be a very crude approximation of the real one. The authors in~\cite{FHIP} try to amend the problem by calculating first order perturbative corrections around the original asantz.

Here we do the opposite. By generalizing Feynman's model, see Fig.~\ref{fig:ModelAbsorption}~B, to a situation where the particle couples linearly with (infinitely) many oscillators, one can readily show~\cite{SBM} that \emph{any} absorption spectrum can be generated by a proper choice of the bath's spectral function. Among them is most definitely the exact absorption spectrum of the polaron. This leaves open an important question: \emph{Can we extract a good approximation for the absorption spectrum by minimizing the ground state energy of the model system ?}

In order to answer this question, the discussion is limited to the case of the optical Fr\"ohlich polaron for which diagrammatic quantum Monte Carlo results are available~\cite{DQMC,Mischenko,Filippis}. Its Hamiltonian, in units of $m=\hbar=1$, is given by
\begin{eqnarray}
H&=&\frac{{\bf p}^2}{2}+\sum_{\bf k} \omega_{k} b^{\dagger}_{\bf k}b_{\bf k}+ \nonumber \\ &+&\sum_{\bf k} \left[ V_k \exp\left(i{\bf k}.{\bf x}\right) b^{\dagger}_{\bf k}+ V^*_k \exp\left(-i{\bf k}.{\bf x}\right) b_{\bf k}\right].
\label{eq:Hamiltonian}
\end{eqnarray}
Any generalized Feynman model gives rise to the following Euclidean action for the reduced density matrix of the particle~\cite{Kleinert}:
\begin{equation*}
\mathcal{S}_0= \int_0^\beta \frac{\dot{\bf x}^2}{2} {\rm d}t-\int_0^\beta\int_0^\beta g\left(t-s\right) ({\bf x}(t) -{\bf x}(s))^2 {\rm d}t{\rm d}s.
\end{equation*}
Here $g(t)$ encodes the spectral properties of the bath. It is useful to define the bath's memory function $\Gamma(\omega)=\omega^2g(\omega)$, with $g(\omega)$ the Fourier transform of $g(t)$. For example, the memory function of Feynman's model is $\Gamma_f(\omega)=(M-1)v^2/(v^2+M\omega^2)$. 

Since the model action $\mathcal{S}_0$ is quadratic, one can analytically find its Euclidean Green function and its free energy. From the Jensen-Feynman inequality
\begin{equation*}
F\leqslant F_0 +\beta^{-1}\left\langle \mathcal{S}-\mathcal{S}_0 \right\rangle,
\end{equation*}
one derives the following bound on the ground state energy of \eqref{eq:Hamiltonian}:
\begin{eqnarray}
E_0&\leqslant& 3 \int_0^{\infty} \frac{{\rm d}\omega}{2\pi} \left[ \log\left(1+\Gamma(\omega)\right)-\frac{\Gamma(\omega)}{1+\Gamma(\omega)}\right]- \nonumber \\ &&-\sum_{\bf k} \left| V_k\right|^2 \int_0^{\infty} {\rm d}t \exp\left(-\omega_k t-\frac{k^2}{2} G(t) \right),
\label{eq:GSEbound}
\end{eqnarray}
with the Euclidean Green function given by
\begin{equation}
G(t)=\int_0^{\infty}{\rm d}\omega \frac{2}{\pi} \frac{1-\cos(\omega t)}{\omega^2(1+\Gamma(\omega))}.
\label{eq:Green}
\end{equation}
Additionally, since the model has a quadratic Hamiltonian, its real time dynamics can trivially be found by solving Heisenberg's equations of motion. The linear response function of the model is completely fixed by the same memory function $\Gamma(\omega)$, i.e. the Laplace transformed conductivity of the generalized Feynman model is
\begin{equation*}
\mathcal{L}(\sigma)(\Omega)=\frac{1}{\Omega+\Omega\Gamma(\Omega)}.
\end{equation*}
Consequently, its optical absorption is simply
\begin{equation*}
\Lambda(\omega)=\lim_{\epsilon\rightarrow 0} {\rm Re} \left[\mathcal{L}(\sigma)(i\omega+\epsilon)\right].
\end{equation*}
There is thus a one-to-one correspondence between the optical absorption spectrum of the model and the variational bound on the ground state energy. Interestingly, the exact ground state energy and optical absorption spectrum are also linked by the ground state energy sum rule introduced in~\cite{LSD}. Remarkably it was pointed out in Ref.~\cite{Thornber1} that all Feynman approximations exactly satisfy this sum rule, even though neither the aborption nor the energy are exact. Unfortunately, these works did not attract much attention. 

Minimization of \eqref{eq:GSEbound} with respect to $\Gamma(\omega)$ results in a non-linear integral equation for the memory function, this integral equation was also obtained by Rosenfelder in~\cite{Rosenfelder}. In general one has to resort to numerical methods to solve it and this seriously hampers the calculation of the optical absorption, as this requires a numerical analytic continuation of the Green's function. Only at weak coupling the energy~\eqref{eq:GSEbound} can be systematically expanded in terms of the electron-phonon coupling strength. 

Indeed, note that one can always rewrite the Green function \eqref{eq:Green} as
\begin{equation}
G(t)=t-\int_0^{\infty}{\rm d}\omega \frac{2}{\pi} \frac{\Gamma(\omega)}{1+\Gamma(\omega)}  \frac{1-\cos(\omega t)}{\omega^2}.
\end{equation}
In the weak coupling regime the second term must be small such that we can expand the exponential in~\eqref{eq:GSEbound} to first order in this term. This yields
\begin{equation*}
E_0\leqslant E_{p}+ 3 \int_0^{\infty} \frac{{\rm d}\omega}{2\pi} \left[ \log\left(1+\Gamma(\omega)\right)-\frac{(1+\Gamma_0(\omega))\Gamma(\omega)}{1+\Gamma(\omega)}\right],
\end{equation*}
where the perturbative energy $E_p$ and the zeroth-order memory function are given by
\begin{eqnarray}
&&E_{p}=\sum_{\bf k} \frac{\left| V_k\right|^2}{\omega_k+k^2/2}, \nonumber \\ &&\Gamma_0(\omega)= \int_0^{\infty} {\rm d}t  \frac{1-\cos(\omega t)}{\omega^2}\sum_{\bf k} \frac{2 k^2\left| V_k\right|^2}{3}e^{-\left(\omega_k+k^2/2\right)t}. \nonumber
\end{eqnarray}
Note that the bound on the energy is preserved by the expansion because of the convexity of the exponential. The energy is minimal for $\Gamma=\Gamma_0$. Hence the weak coupling expansion of the energy, to order $V_k^4$, becomes
\begin{equation*}
E_0\leqslant E_{p}- \frac{3}{2} \int_0^{\infty} \frac{{\rm d}\omega}{2\pi} \Gamma_0(\omega)^2.
\end{equation*}
For the optical Fr\"ohlich polaron the electron-phonon coupling is given by $\left| V_k\right|^2=2\sqrt{2}\pi \alpha/Vk^2$, where $\alpha$ is the dimensionless coupling strength. In the continuum limit $E_p=-\alpha$ and 
\begin{equation}
\Gamma(\omega)=\frac{4\alpha}{3\omega^2} \left(\frac{\sqrt{\sqrt{\omega^2+1}+1}}{\sqrt{2}}-1 \right).
\label{eq:weakgamma}
\end{equation}
The numerical value of the weak coupling bound on the energy is therefore
\begin{equation*}
E_0\approx-\alpha-1.25978 \left( \frac{\alpha}{10} \right)^2.
\end{equation*}
The $\alpha^2$ coefficient ought to be compared to Feynman's $1.23$ and to the exact weak coupling result $1.592$~\cite{Hohler,Roseler}. The present result is clearly lower than Feynman's but the improvement is very small. Moreover, we do not recover the exact result. The gain in energy, however small, results in a drastic change of the memory function, i.e. from Feynman's asatz $\Gamma_f$ to expression \eqref{eq:weakgamma}. 
Despite the fact that both functions have similar low frequency behaviour, they differ significantly at high frequency. Where Feynman's memory function decays quadratically, the optimal weak coupling memory function only decays like $\omega^{-3/2}$. Even though this seems to change only little to the behaviour of the function on the real axis it drastically alters its properties in the complex plane, resulting in a completely different optical absorption.

\begin{figure}[h]
\centering
\includegraphics[width=0.5\textwidth]{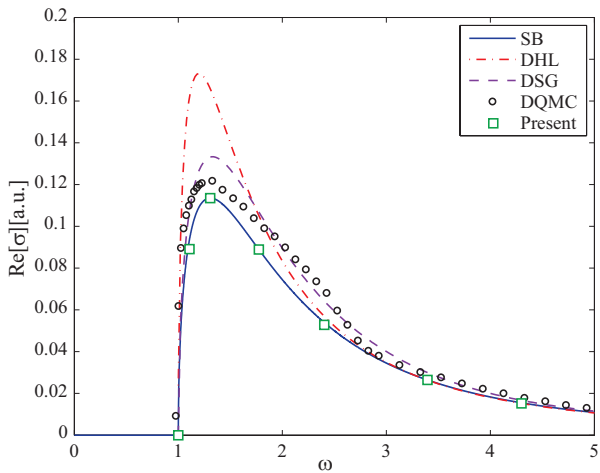}
\caption{(Color Online) Optical absorption coefficient for Fr\"ohlich polaron at $T$=$0$ for $\alpha$=$1$. The full blue line (SB) represent the author's recent result based on a truncated Wigner approximation~\cite{SBdynamic}, the dashed red line (DHL) is a perturbative result by Devreese et al.~\cite{DHL}. Furthermore, the dashed purple line (DSG) is a variational result due to Devreese et al.~\cite{DSG} and the circles (DQMC) show a numerical result due to Mishchenko et al.\cite{Mischenko}. The green squares indicate the present result.}
\label{fig:alpha1abs}
\end{figure}

The optical absorption is compared with previous results and quantum Monte Carlo data for $\alpha=1$ in Fig.~\ref{fig:alpha1abs}. The present absorption clearly is much more realistic than the model absorption in FHIP and is comparable to DSG. Furthermore note that the absorption is mathematically identical to the author's previous result obtained by rigorously truncating kinetic equations in the Weyl representation~\cite{SBmobility,SBdynamic}. Given the drastically different nature of both descriptions this is remarkable. Note that for $\alpha=1$ one already expects deviations from the perturbative result and the actual self consistent absorption should even be closer to the Monte Carlo data. When the coupling is truly small, i.e. $\alpha \lesssim 0.1$, all the optical conductivities (DHL,DSG,SB) coincide.

Although the previous bound on the energy is rigorous, it becomes rather weak at strong coupling where the expansion of the exponential in expression~\eqref{eq:GSEbound} becomes problematic. At strong coupling one must resort to numerical methods to find the full self-consistent Green function. The numeric problem was previously considered in the works~\cite{Saitoh, Rosenfelder}. The present analytic weak coupling result confirms their findings, i.e. we exactly reproduce the $1.25978\times 10^{-2} \alpha^2$ correction to the weak coupling energy. 

\begin{figure}[tp]
\centering
\includegraphics[width=0.5\textwidth]{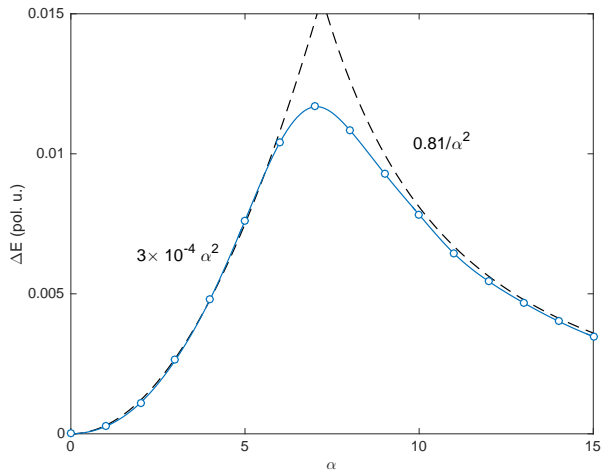}
\caption{(Color Online) The energy difference, in polaronic units, between the ground state energy obtained by Feynman and by the generalized Feynman model. The difference grows quadratically at weak coupling. At strong coupling the difference decays like $\alpha^{-2}$. }
\label{fig:energydiff}
\end{figure}

Given how close Feynman's result is to the exact energy, there is little to gain. Numerical results indeed show that at strong coupling there is no change in the leading order $\alpha^2$ scaling, and corrections to the ground state energy vanish like $\alpha^{-2}$ in agreement with Refs.~\cite{Saitoh, Rosenfelder}. The difference in ground state energy between the full variational model and Feynman's model is depicted in Fig.~\ref{fig:energydiff}. Note that this is consistent with adiabatic strong-coupling expansions, which can only change the leading order term by resorting to non-Gaussian wave functions~\cite{LP,Miyake}.

Weak coupling results have shown that a very small change in the energy might however be accompanied with a rather large change in the dynamic response of the system, but would a vanishing correction of $\mathcal{O} \left(\alpha^{-2}\right)$ be sufficient to give the absorption peak a finite width? The Supplementary Information describes a simplified, analytically tractable, model which shows it is. The model contains an explicit parameter for the peak width but it has infinite effective mass. The models memory kernel is simply
\begin{equation*}
\Gamma(\omega)=\frac{\omega_0^2}{\omega^2}+\frac{\delta}{\omega}
\end{equation*}
 At large coupling, the width $\delta$ of the optimal model tends to a constant like $\alpha^{-2}$. Consequently, up to a constant, the width is proportional to the inverse of the Franck-Condon frequency as discussed in \cite{KED}. Moreover, the correction to the ground state energy is only of $\mathcal{O} \left(\alpha^{-2}\right)$. A clear indication that the fully self consistent numeric result should have a finite width for the absorption peak in the large coupling limit.

A detailed comparison between the numerical and Feynman's memory function furthermore reveals that the low frequency behaviour of the optimal memory function $\Gamma(\omega)$ is completely consistent with Feynman's. However, at high frequency both memory functions start to deviate. The optimal $\Gamma$ falls of slower then $\omega^2$, i.e. the scaling is consistent with a $-3/2$ power law, similar to the high frequency scaling at weak coupling. Note that the same asymptotic scaling is also predicted for the optical conductivity from DSG~\cite{DSG}. 

\begin{figure}[bp]
\centering
\includegraphics[width=0.51\textwidth]{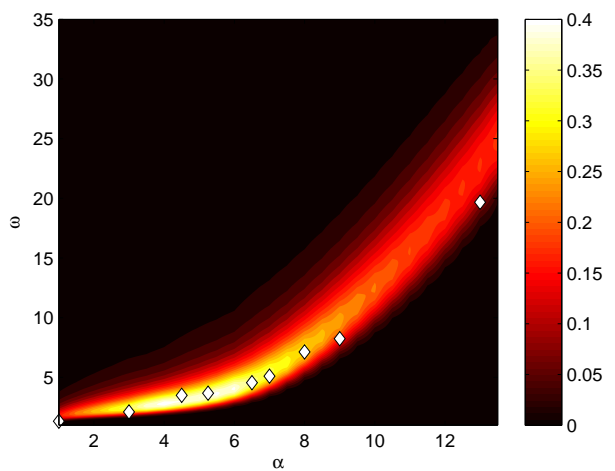}
\caption{(Color Online) Numerical estimate of the optical absorption spectrum of the optimal generalized Feynman model. White diamonds indicate the point of maximum absorption in the DQMC data.}
\label{fig:absumeric}
\end{figure}

Finally we extract the optical conductivity from the numeric data by numeric analytic continuation. In the present formulation this would require the inversion of a Hilbert matrix. Unfortunately this is the canonical example of an ill-conditioned problem. In order to stabilize the analytic continuation one must reduce the number of degrees of freedom, details on this can be found in the supplementary information. Note that this is just another way to see that the free energy is not sensitive to detailed dynamic properties. The resulting absorption is depicted in Fig.~\ref{fig:absumeric}.  

In weak to intermediate coupling, the absorption is in good agreement with Monte Carlo results. At strong coupling the variational results are less accurate. Allthough the violation of the Heisenberg uncertainty principle is resolved, the absorption peaks are still to narrow. In contrast to Ref.~\cite{DSG}, which shows excellent agreement of the peak position with DQMC data, the present spectra have maximum absorption at slightly to high frequency. The latter is consistent with the strong coupling adiabatic expansion using Gaussian wave functions~\cite{Klimin}. At strong coupling there are quantum fluctuations which can not be captured to first-order by any quadratic (Gaussian) model.  

In conclusion, there always exists a quadratic generalized Feynman Lagrangian that has the same optical conductivity as the real system. It follows from Thornbers arguments in~\cite{Thornber1} that the Jensen-Feynman inequality can not be used to find this \emph{exact} absorption spectrum. Nevertheless, decent agreement between the DQMC absorption and the optimal Feynman approximation is found. However, the main message of this paper might be one of caution. No matter how close you get the free energy of a model calculation to the exact result, one might completely miss the actual physics of the problem.
  
At present the model Lagrangian does give the dynamical behaviour of the system and so, in contrast to the approach of FHIP, the results are accurate without perturbative corrections. Finally it was shown that the weak coupling result is identical to that in~\cite{SBdynamic}, providing indirect evidence for the correctness of those results.

Given the simplicity of the present method it seems likely that this can be combined with quantum Monte Carlo calculations to improve their convergence. Moreover, since the dynamics of quantum systems with quadratic actions is identical to their classical dynamics, a diagrammatic expansion around the optimal quadratic Lagrangian would truly separate quantum events from the effective classical dynamics. Further extension might follow the lines of Feynman and Kleinert~\cite{FK,Kleinert}.

\acknowledgments
I would like to thank J.T. Devreese for many discussions on this topic and for his detailed comments on the manuscript. I acknowledge discussions with F. Brosens and thank him for encouraging me to publish my findings. I also thank J. Tempere, A. Polkovnikov, S. Klimin for valuable discussions.
I am financially supported by the FWO as post-doctoral fellow of the Research Foundation - Flanders.

\newpage
\onecolumngrid
\appendix
\section{Supplementary Information: \\
Dynamic polaron response from variational imaginary time evolution}

\section{Analytic strong coupling model}
In order to confirm that the best variational model does not have an optical absorption that tends to a delta function at strong coupling I propose the following ansatz for $\Gamma$:
\begin{equation*}
\Gamma(\omega)=\frac{\omega_0^2}{\omega^2}+\frac{\delta}{\omega}.
\end{equation*}
The resulting optical conductivity will have a peak at $\omega=\omega_0$ and the width of the peak will be $\delta$. Because of the $1/\omega$ term the model energy diverges, however the expected model influence phase also diverges so that the total contribution of the model is finite:
\begin{equation*}
E_m= 3 \int_0^{\infty} \frac{{\rm d}\omega}{2\pi} \left[ \log\left(1+\Gamma(\omega)\right)-\frac{\Gamma(\omega)}{1+\Gamma(\omega)}\right]=\frac{3}{2\pi} \left(\delta +\frac{\omega_0}{\sqrt{1-(\delta/2\omega_0)^2}}\left(\frac{\pi}{2}-\arctan\left(\frac{\delta/2\omega_0}{\sqrt{1-(\delta/2\omega_0)^2}}\right) \right) \right),
\end{equation*}
whenever $\delta>0$. We are only interested in the strong coupling limit which implies $\omega_0\rightarrow\infty$. So asymptotically we find
$E_m \approx \frac{3\omega_0}{4}+\frac{3\delta}{4\pi}+\frac{3\delta^2}{32\omega_0}$. Next we need the contribution of the self energy. Let us, similar to the weak coupling expansion, expand both $G$ and the exponential around $\delta=0$, i.e. we rewrite the Green function as
\begin{equation*}
G(t)=\int_0^{\infty}{\rm d}\omega \frac{2}{\pi} \frac{1-\cos(\omega t)}{\omega^2+\omega_0^2}- \int_0^{\infty}{\rm d}\omega  \frac{2\omega\delta}{\pi} \frac{1-\cos(\omega t)}{(\omega^2+\omega_0^2)(\omega^2+\omega_0^2+\delta \omega) }= G_0(t)- G_1(t),
\end{equation*}
and expand the result in $G_1$. The zeroth order Green function is given by
\begin{equation*}
G_0(t)=\frac{1}{\omega_0} \left(1-\exp\left(-\omega_0 t \right)\right).
\end{equation*}
Hence it's contribution to the self-energy is
\begin{equation*}
\Sigma_0=\frac{\sqrt{2}\alpha}{\pi}\int_0^\infty {\rm d}k \int_0^{\infty} {\rm d}t \exp\left(-t-\frac{k^2}{2} G_0(t) \right)=\frac{\alpha}{\sqrt{\omega_0}} \frac{\Gamma\left(\frac{1}{\omega_0} \right)}{\Gamma\left(\frac{1}{2}+\frac{1}{\omega_0} \right)}, 
\end{equation*}
where $\Gamma(x)$ denotes the Gamma function. For large $\omega_0$ this becomes
\begin{equation*}
\Sigma_0=\frac{\alpha}{\sqrt{\pi}} \left( \sqrt{\omega_0}+ \frac{2 \log 2}{\sqrt{\omega_0}} +\mathcal{O} \left( \omega_0^{-3/2}\right)\right)
\end{equation*}
The first order correction is given by
\begin{equation*}
\Sigma_1=\frac{\alpha\delta}{\sqrt{2}\pi}\int_0^\infty {\rm d}k k^2 \int_0^{\infty} {\rm d}t \,G_1(t) \exp\left(-t-\frac{k^2}{2} G_0(t) \right)= \frac{\alpha\delta}{2\sqrt{\pi}} \int_0^{\infty} {\rm d}t \frac{G_1(t)}{ G_0^{3/2}(t)} \exp\left(-t\right). 
\end{equation*}
It seems like this integral can not be done analytically in general. At strong coupling one can however make some approximations. For sufficiently large $\omega_0$ one can approximate $G_0=1/\omega_0$. Next all integrals can be done. Systematically keeping only terms to order $\omega_0^{-1/2}$ yields only a single contribution, i.e.
\begin{equation*}
\Sigma_1= \frac{\alpha \delta}{2\pi\sqrt{\pi\omega_0}}+\mathcal{O} \left( \omega_0^{-3/2}\right).
\end{equation*}
The total energy thus becomes
\begin{equation}
E_0=\frac{3\omega_0}{4}-\frac{\alpha}{\sqrt{\pi}}\left(\sqrt{\omega_0}+\frac{2\log(2)}{\sqrt{\omega_0}}\right)+\frac{3\delta}{4\pi}\left(1-\frac{2\alpha}{3\sqrt{\pi\omega_0}}\right)+\frac{3\delta^2}{32\omega_0}.
\label{eq:GSa}
\end{equation}
There will thus be a finite value of the width $\delta$ if $\left(1-\frac{2\alpha}{3\sqrt{\pi\omega_0}}\right)<0.$ Optimizing with respect to $\delta$ we have
\begin{equation*}
\delta=\frac{4\omega_0}{\pi}\left(\frac{2\alpha}{3\sqrt{\pi\omega_0}}-1\right).
\end{equation*}
Substituting this back into expression \eqref{eq:GSa} and optimizing with respect to $\omega_0$ yields \begin{equation*}
\omega_0=\frac{4\alpha^2}{9\pi}-4\log(2)\frac{\pi^2}{\pi^2-2}+\mathcal{O}(\alpha^{-2}).
\end{equation*}
Consequently the width behaves as \begin{equation*}
\delta=\frac{8\log(2)}{\pi}\frac{\pi^2}{\pi^2-2}+\mathcal{O}(\alpha^{-2})
\end{equation*}
Note that the variational energy only changes to $\mathcal{O}(\alpha^{-2})$. Clearly the optimal width in this very simple model already tends to a finite value at large $\alpha$ and not to zero. Note that this is only possible because  $\left(1-\frac{2\alpha}{3\sqrt{\pi\omega_0}}\right)<0$. If we were to do an adiabatic calculation, i.e. calculate $\omega_0$ from the Landau-Pekar ansatz, then $\omega_0=4\alpha^2/9\pi$. Hence within the adiabatic (Born-Oppenheimer) approximation the frequency would make the linear term in $\delta$ vanish so that the optimal linewidth $\delta=0$. Incorporating non-adiabatic effects results in slightly lower excitation frequency $\omega_0$. The linewidth is directly proportional to the difference between the adiabatic and the real excitation frequency. The result clearly shows the relation between dissipation and non-adiabaticity.

\section{Optical conductivity numerics}
By analytically continuating the Euclidian Green function we find the following expression for it in terms of the optical absorption
\begin{equation*}
G(t)=\int_0^t \mathrm{d}s \int_0^{\infty} \frac{\mathrm{d}\omega}{\pi}  \Lambda(\omega) \exp(-\omega s).
\end{equation*}
Since we have numerically calculated the left hand side, finding $\Lambda$ is just a matter of solving a linear equation. Before I go on let  me just show that this is numerically ill defined. Solving a linear equation is formally identical to minimizing the total square of the error. It's instructive to minimize the error on the derivative of $G$ rather then on $G$ itself. Minimizing the error
\begin{equation*}
D_1=\int_0^{\infty} \mathrm{d}t \left(\partial_t G(t)-\int_0^{\infty}  \frac{\mathrm{d}\omega}{\pi} \Lambda(\omega) \exp(-\omega t) \right)^2,
\end{equation*}
with respect to $\Lambda$ yields
\begin{equation*}
\int_0^\infty \frac{\mathrm{d}\nu}{\pi} \frac{\Lambda(\nu)}{\nu+\omega}=\omega G(\omega).
\end{equation*}
Here $G(\omega)$ is the Laplace transform of $G(t)$. When discretized on an equidistant grid the above problem results in the problem of inverting a Hilbert matrix. The determinant of this matrix asymptotically scales like $4^{-N^2}$, where $N$ is the number of grid point. The latter implies eigenvalues are at least exponentially small in the number of grid points. If we wish to have any accuracy we need to numerically know the Green function to the same (exponential) precision.
Note that this is just another way of observing why the ground state energy is not sensitive to detailed dynamical properties. The exponentially small eigenvalues imply that a lot of $\Lambda$ will give a very good fit to $G$, in the sense that $D_1$ will be very small. It thus becomes a matter of selecting the physical solution out of a set of, to numerical accuracy, degenerate solutions. 
In order to solve this problem one generally resorts to Maximum entropy methods~\cite{Silver}. The idea is to introduce a penalty for solutions that are to far away from a (smooth) prior distribution. This introduces an extra parameter into the problem. If one regards the distance $D$ as the energy one should in principle look for the ground state of this but in the maximum entropy method (MEM) you look for states that minimize free energy. The temperature is the extra degree of freedom.  

\begin{figure}[h]
\centering
\includegraphics[width=0.45\textwidth]{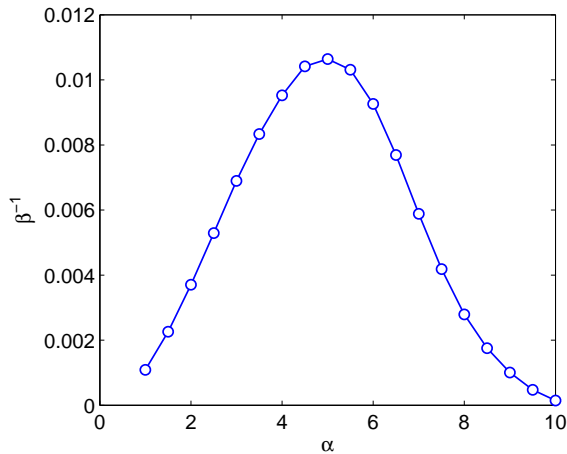}
\caption{The temperature in the numeric analytic continuation scheme for the optical Fro\"hlich polaron as a function of the coupling constant. This serves as an indirect indication of the numerical error.}
\label{fig:temp}
\end{figure}

Here I obtained the absorption in the following way. First of all I subtract the linear contribution to the Green function, as this results in a delta function at $\omega=0$. In fact I fix the effective mass to $m^*/m=1+\Lambda(0)$ and only numerically look for $\Lambda$ at finite frequency. Next I define the distance  \begin{equation*}
D(\omega_0)=\int_0^{\infty} \mathrm{d}t \left( G(t)-\int_0^\infty\mathrm{d}s\int_0^{\infty}  \frac{\mathrm{d}\omega}{\pi} \Lambda_0(\omega) \exp(-\omega s) \right)^2,
\end{equation*}
where $\Lambda_0=\pi/2\left(\delta(\omega)/m^*+(m^*-1)/m^*f(\omega-\omega_0)\right)$ and I take $f(\omega)$ to be a normal distribution with a fixed width, at present its variance is 1. In this way the ansatz $\Lambda_0$ satisfies the f-sum rule and its resolution is low compared to the underlying grid. Minimizing $D(\omega_0)$ with respect to $\omega_0$ results in the best Gaussian fit to the absorption spectrum. 
However, at this point one can argue that similar to MEM one can take a thermal state of $D$ rather then the ground state. Therefore I write
$\Lambda(\omega)=Z^{-1}\exp(-\beta D(\omega)).$ The partition function can be fixed by the f-sum rule. Furthermore the temperature can be fixed by the ground state energy sum rule~\cite{LSD}. Note that the temperature is now a direct measure for the quality of the fit. As shown in Fig.~\ref{fig:temp}, the temperature reaches a maximum in the intermediate coupling regime. The present numerical results should thus be most accurate at weak and strong coupling. More sophisticated schemes are required to accurately determine the response at intermediate coupling.

\end{document}